\documentclass[twocolumn,showpacs,amssymb]{revtex4}
\usepackage{graphicx}
\usepackage{psfrag}
\usepackage{amsmath}
\newcommand{\ket}[1]{\mbox{$\left| #1 \right\rangle $}}
\newcommand{\bra}[1]{\mbox{$\left\langle #1 \right| $}}
\newcommand{\ps}[2]{\mbox{$\left\langle #1 |#2 \right\rangle $}}
\begin{document}

\title{Implementation of STIRAP in degenerate systems by dimensionality reduction}

\author{G. Bevilacqua$^{1}$, G. Schaller$^{2}$, T. Brandes$^{2}$,  and F. Renzoni$^{3}$}

\affiliation{$^{1}$Department of Information Engineering and Mathematical Science, University of Siena, Via Roma 56, 53100 Siena, Italy}
\affiliation{$^{2}$Institut f\"ur Theoretische Physik, Technische Universit\"at Berlin, Hardenbergstrasse 36, 10623 Berlin, Germany}
\affiliation{$^{3}$Department of Physics and Astronomy, University College London, Gower Street, London WC1E 6BT, United Kingdom}
\date{\today}

\begin{abstract}
We consider the problem of the implementation of Stimulated Raman Adiabatic 
Passage (STIRAP) processes in degenerate systems, with a view to be able to steer
the system wave function from an arbitrary initial superposition to an arbitrary target 
superposition. We examine the case a $N$-level atomic system consisting of $ N-1$ 
ground states coupled to a common excited state by laser pulses. We analyze the general
case of initial and 
final superpositions belonging to the same manifold of states, and we cover also the case 
in which they are non-orthogonal. We demonstrate that, 
for a given initial and target superposition,  it is always possible to choose the laser pulses 
so that in a transformed basis the system is reduced to an effective three-level $\Lambda$ 
system, and standard STIRAP processes can be implemented. Our treatment leads to a  
simple strategy, with minimal computational complexity, which allows us to determine  the 
laser pulses shape required for the wanted adiabatic steering.
\end{abstract}


\maketitle

\section{Introduction}

Destructive quantum interference allows the control of the properties of quantum systems
as well as their evolution in time. Many important features can be understood by 
considering a three-level atom consisting of two ground states coupled to a common 
excited state by two laser fields. Whenever the detuning between the two laser fields
matches the ground state splitting, the system is prepared into a superposition of the
ground state which is decoupled from the laser radiation - the so called {\it dark state}
\cite{moi, ari76,ari}. This also allows the control of the absorptive and dispersive 
properties of a medium consisting of three-level atoms \cite{harris_rev}.

Additional interesting features appear for time dependent laser fields. In this case 
the dark state becomes time-dependent, and this allows one to control the quantum 
state of the atom via adiabatic following of the dark state - the so called Stimulated 
Raman Adiabatic Passage (STIRAP) \cite{rmp98}. STIRAP is not directly applicable
to degenerate systems as in this case the system may have several 
dark states, so that the non-adiabatic coupling between them is not negligible
and the adiabatic theorem does not directly apply. Several strategies have been
developed for the adiabatic steering of degenerate quantum systems in 
different configurations. This led to a number of schemes for the creation and
manipulation of superpositions \cite{shore95,martin95,kis01,kis02,kis04,boozer,shapiro}, 
as well as schemes for the implementation of quantum gates based on STIRAP
\cite{renzoni02,goto04,jauslin05}.  Of particular relevance for the work presented here,
is previous work dealing with an atomic system consisting of a multiplet of 
degenerate ground states coupled to a common excited state by laser pulses. 
Solutions for the steering of arbitrary superposition were identified by using
numerical optimal control techniques \cite{kis02}. Analytic solutions were also 
found for specific configurations \cite{kis01}. Analytic solutions of the nondegenerate 
quantum control problem in the case of arbitrary initial and final superpositions 
belonging to different manifold of states were given in Ref. \cite{shapiro}.

In this work we consider the problem of steering the atomic wave function by 
STIRAP in an $N$-level atomic system consisting of $ N-1$ ground states coupled
to a common excited state by laser pulses. We analyze the general case of initial and 
final superpositions belonging to the same manifold of states, and we cover also the case 
in which they are non-orthogonal. We demonstrate that, for a given 
initial and target superposition,  it is always possible to choose the laser pulses so 
that in a transformed basis the system is reduced to an effective three-level system, 
and standard STIRAP processes can be implemented. Our treatment leads to a 
simple strategy, with minimal computational complexity, which allows us to determine 
the laser pulse shapes required for the wanted adiabatic steering.

This work is organized as follows. 
In Sec. \ref{sec:statement} we define the system of interest, and state 
the problem under consideration. In Sec. \ref{sec:theory} we derive the conditions for the 
reduction of the system to an effective three-level $\Lambda$ system. 
We then specify the conditions on the laser pulses for the transfer from a
given initial superposition to a wanted final superposition. 
In Sec. \ref{sec:numerics} we demonstrate the validity of our approach with 
numerical simulations. Conclusions are drawn in Sec. \ref{sec:conclusions}.

\section{Statement of the Problem}\label{sec:statement}

We consider an $N$ level atomic system with  $N-1$  degenerate ground states
coupled to a common excited state $|N\rangle$ by laser fields of equal frequency 
$\omega$, taken to be equal to the atomic transition frequency.
 This is the same model considered in Refs. \cite{kis02,kis01} to understand the 
mechanism of STIRAP processes in systems with a degenerate dark state subspace.
The scheme finds direct application in the creation and manipulation of atomic systems.
For $N=4$ it directly describes an atomic system with three degenerate ground states 
coupled to a common excited state by fields of different polarizations. The procedure 
identified in this work also applies, for larger $N$,  to level schemes including non 
degenerate ground state sublevels, e.g.  sublevels of different hyperfine states. In 
this case each  level is individually resonantly coupled to a common excited state by 
a laser field of appropriate frequency and polarization. This gives rise to  a 
degenerate dark space for which the procedure of implementation of STIRAP
identified in this work applies.

\begin{figure}[ht]
  \begin{center}
  \includegraphics[width=3in]{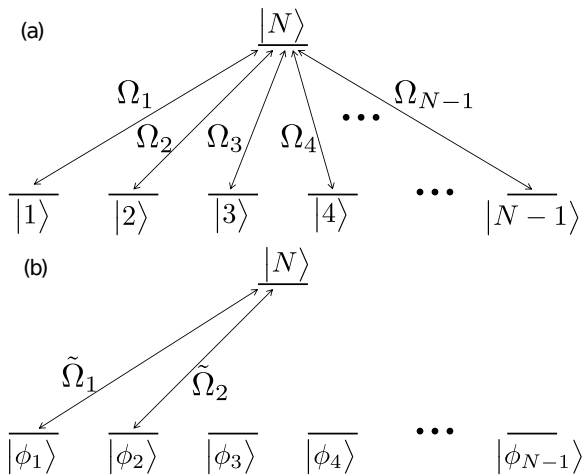}
\end{center}
  \caption{$N$-level system consisting of $N-1$ ground states coupled to a common
excited state. (a) Interaction scheme in the basis $\{|i\rangle\}$ with couplings $\Omega_i$, 
as read from Eq. \eqref{eq:Ham:ini}
   (b) Interaction    scheme in the basis $\{|\phi_i\rangle\}$  with couplings $\tilde{\Omega}_i$,
 as from  Eqs.  \eqref{eq:cond:W}.}
  \label{fig:lev}
\end{figure}

 In a frame 
rotating at frequency $\omega$, the Hamiltonian in the rotating-wave 
approximation (RWA) can be written as
\begin{equation}
  \label{eq:Ham:ini}
  H = \sum_{i=1}^{N-1} \, \hbar\Omega_i(t) \bigg[ \ket{i}\bra{N} + \ket{N}\bra{i} \bigg]~,
\end{equation}
where $\Omega_i (t)$ is the time-dependent Rabi frequency for the 
transition $|i\rangle \to |N\rangle$. The Rabi frequencies are taken as real without
loss of generality, as any complex phase can be re-absorbed into a re-definition of 
the basis states. The interaction scheme is represented in Fig. \ref{fig:lev}. The 
system has a subspace of superposition of ground states decoupled from the laser 
fields ("dark state subspace") of dimension $N-2$ \cite{kis02}.

We aim to determine a set of laser pulses  $\Omega_i(t)$  which drives the atomic 
system from an arbitrary initial ground state superposition $|\psi_i\rangle$
\begin{equation}
  \label{eq:psi:ini}
  \ket{\psi_i}  =\sum_{i=1}^{N-1} x_i \, \ket{i}. 
\end{equation}
to an arbitrary final target ground state superposition $|\psi_f\rangle$
\begin{equation}
  \label{eq:psi:fin}
  \ket{\psi_f} = \sum_{i=1}^{N-1} y_i \, \ket{i}. 
\end{equation}
We restrict our analysis to the case of temporal evolution determined
by the adiabatic following of a  dark state, without any 
mixing with the excited state. 

\section{Theoretical analysis}\label{sec:theory}

For clarity, we consider separately the two cases of  orthogonal and
non-orthogonal initial and target states. We first discuss the orthogonal 
case in Sec. III.A while the general case is discussed in Sec. III.B

\subsection{Case I:  orthogonal initial and target states}

We consider the case of  orthogonal initial and target states 
\begin{equation}
  \label{eq:ortho}
  \ps{\psi_f }{ \psi_i} = 0~.
\end{equation}
We introduce a new atomic basis for the ground state subspace in which the 
first two states are the initial and the target states $\ket{\psi_i}$ and $\ket{\psi_f}$. 
The basis is then completed by $N-3$ linear combinations of the original ground states, 
as it can be obtained by standard Gram - Schmidt orthogonalization procedure: 
\begin{align}
  \label{eq:orto:lin:com}
  \ket{\phi_1} & = \ket{\psi_i}\,, \\
  \ket{\phi_2} & = \ket{\psi_f}\,, \\
  \ket{\phi_3} & =  \sum_i \alpha_3^i \ket{i}\,, \\
  \vdots &  \\
  \ket{\phi_{N-1}} & = \sum_i \alpha_{N-1}^i \ket{i}~,
\end{align}
where the coefficients $\alpha_j^i$ are determined by the orthogonalization procedure.
For notational convenience we rewrite this as
\begin{equation}
  \label{eq:comb:lin:matr}
  \ket{\phi_i} = \sum_j C_{ij} \ket{j} \qquad i,j=1,\ldots,N-1~.
\end{equation}
The  matrix  $C$  is  orthogonal   and  the  first  two  rows
correspond to the coefficients $x_i$ and $y_i$ of the initial 
and the target superpositions, respectively. 

In the new basis the Hamiltonian reads as
\begin{equation}
  \label{eq:H:new}
  H = \sum_{i=1}^{N-1} \, \tilde{\Omega}_i(t) \bigg[ \ket{\phi_i}\bra{N} + \ket{N}\bra{\phi_i} \bigg]
\end{equation}
where the transformed  pulses are defined as
\begin{equation}
  \label{eq:new:pulses}
  \tilde{\Omega}_i(t) =  \sum_j C_{ij} \Omega_j(t) ~.
\end{equation}
Notice how the  transformed Hamiltonian has the same  structure of the
initial one. We  can thus easily find the  dark states associated with
(\ref{eq:H:new}). We first parametrize the $N-1$ laser pulses $\tilde{\Omega}_i$
in terms of a total amplitude $\Omega$
\begin{equation}
\Omega = \left( \sum_{i=1}^{N-1} \tilde{\Omega}_i^2 \right)^{1/2}
\label{eq:totalomega}
\end{equation}
and $N-2$ angles $\theta_1,...,\theta_{N-2}$
as
\begin{align}
  \label{eq:para:pulse}
  \tilde{\Omega}_1 & = \Omega\sin \theta_{N-2} \; \sin \theta_{N-3 }\; \cdots\;
  \sin \theta_2 \; \sin \theta_1~, \\
  \tilde{\Omega}_2 & = \Omega\sin \theta_{N-2} \; \sin \theta_{N-3 }\; \cdots\;
  \sin \theta_2 \; \cos \theta_1~, \\
  \tilde{\Omega}_3 & = \Omega\sin \theta_{N-2} \; \sin \theta_{N-3 }\; \cdots\;
  \cos \theta_2 ~, \\
  \vdots \\
  \tilde{\Omega}_{N-3} &  = \Omega \sin \theta_{N-2} \;  \sin \theta_{N-3 }
  \; \cos \theta_{N-4}~, \\
  \tilde{\Omega}_{N-2} & = \Omega \sin \theta_{N-2} \; \cos \theta_{N-3 }~,\\
  \tilde{\Omega}_{N-1} & = \Omega \cos \theta_{N-2}~.
\end{align}
By forming the state 
\begin{equation}
  \label{eq:dark:0}
 |\chi\rangle = \sum_j \tilde{\Omega}_j |\phi_j\rangle~,
\end{equation}
the $N-2$ dark states $\ket{\chi_k}$ can be  easily obtained (apart a normalization factor)
as \cite{kis01}:
\begin{equation}
\ket{\chi_k}  = \frac{\partial  }{\partial \theta_k}  \ket{\chi} ~.
\label{eq:darkk}
\end{equation}
Of particular relevance for the following is the first dark state, which reads
\begin{equation}
 \ket{\chi_1}   = \frac{\partial  }{\partial \theta_1}  \ket{\chi} \propto \cos \theta_1 \ket{\phi_1} - \sin \theta_1 \ket{\phi_2}~.
\label{eq:dark1}
\end{equation}

% \begin{align*}
 %  \ket{\chi_1}  & = \frac{\partial  }{\partial \theta_1}  \ket{\chi} =
 % \cos \theta_1 \ket{\phi_1} - \sin \theta_1 \ket{\phi_2}\\
 % \ket{\chi_2}  & = \frac{\partial  }{\partial \theta_2}  \ket{\chi} =
 % \cos  \theta_2  \sin  \theta_1  \ket{\phi_1} +  \cos  \theta_1  \cos
 % \theta_2 \ket{\phi_2} - \sin \theta_2 \ket{\phi_3}\\
 % \vdots &
%\end{align*}

In the basis $|\phi_i\rangle$ with couplings $\tilde{\Omega}_i$ it is 
immediate to see that the system can be reduced by an appropriate 
choice of the laser pulses to a three-level $\Lambda$ system 
consisting of the states $\{ \ket{ \psi_i}, \ket{ \psi_f }, \ket{N }\}$, so 
that adiabatic transfer from $ \ket{ \psi_i}$ to $ \ket{ \psi_f }$ can 
be implemented. 
%%GS BEGIN

In this subspace, the energy eigenstate corresponding to a eigenvalue $\lambda=0$ is the first
dark state $\ket{\Psi_0} \propto \tilde{\Omega}_2(t) \ket{\phi_1} - \tilde{\Omega}_1(t) \ket{\phi_2}$.
A sufficient condition to remain in this energy eigenstate throughout the evolution is (see e.g. Eq.~(6) 
in~\cite{schaller2006b})
\begin{equation}\label{Econdition_adiabatic}
\left|\frac{\tilde{\Omega}_1(t) \dot{\tilde{\Omega}}_2(t) - \tilde{\Omega}_2(t) \dot{\tilde{\Omega}}_1(t)}
{\sqrt{2} \left[\tilde{\Omega}_1^2(t)+\tilde{\Omega}_2^2(t)\right]^{3/2}}\right| \ll 1\qquad\forall t\,,
\end{equation}
and may now serve to find optimized control functions $\tilde{\Omega}_{1/2}(t)$ with the side constraints
that initially $\tilde{\Omega}_2 \gg \tilde{\Omega}_1$ and finally $\tilde{\Omega}_1 \gg \tilde{\Omega}_2$.
It should be noted that fully adiabatic evolution may be a rather strict criterion, as it is for 
practical application only required that the final state of the system corresponds to the final dark state.
In situations where the adiabatic preparation time is an issue, it may therefore be favorable to find more optimal
control functions minimizing only the final occupations of the other two eigenstates.

Taking further into account typical experimental implementations, it is convenient to choose the
transformed  Rabi frequencies as
%%GS END
\begin{subequations}
  \label{eq:cond:W}
  \begin{align}
    \tilde{\Omega}_1 & \equiv f(t) \label{eq:cond:W1}~,\\
    \tilde{\Omega}_2 & \equiv f(t+\tau) \label{eq:condW:2}~,\\
    \tilde{\Omega}_j & \equiv 0 \qquad j > 2 \label{eq:condW:3}~,
\end{align}
\end{subequations}
where $f(t)$ has to be compatible with the conditions above.
%GS the exact conditions for $f(t)$ will be specified later. 
%
As it will be shown, this choice leads to physical laser pulses which are linear combinations of 
delayed pulses, and are easy to implement.
Condition
(\ref{eq:condW:3}) determines the reduction of the system to 
an effective three-level $\Lambda$ system, with the two states $ \ket{ \phi_1},  \ket{ \phi_2}$
(i.e. $\ket{ \psi_i} , \ket{ \psi_f})$ coupled via a common excited state.
The remaining $N-3$ states are spectator ground states not involved in the process.  
The reduction to an effective $\Lambda$ system for an appropriate choice of 
laser pulses is shown in Fig.~\ref{fig:lev}(b).

We can then implement a standard STIRAP process by taking a pair of pulses
$f(t), f(t+\tau)$ which satisfy the standard requirements of STIRAP in terms of 
smoothness, duration, strength and overlap. The system has a dark state, Eq. (\ref{eq:dark1}).
The temporal dependence of the angle $\theta_1$ is determined by the temporal-dependence
of $\tilde{\Omega}_1$, $\tilde{\Omega}_2$. In the specific case
\begin{equation}
\tan \theta_1 = \frac{\tilde{\Omega}_1}{\tilde{\Omega}_2}=\frac{f(t)}{f(t+\tau)}~,
\end{equation}
which implies that $\theta_1 (t\to -\infty)=0$ and  $\theta_1 (t\to +\infty)=\pi/2$. The 
dark state has thus the properties 
\begin{align}
  \label{eq:chi1}
  \ket{\chi_1(-\infty) } &= \ket{ \psi_i }~, \\
  \ket{\chi_1(+\infty) } &= \ket{ \psi_f } ~.
\end{align}
Therefore as in standard STIRAP in a three-level system, adiabatic evolution along
the dark state  $\ket{\chi_1} $ will lead to the transfer of the system from
$\ket{\psi_i}$ to $\ket{\psi_f} $.

The  conditions  \eqref{eq:cond:W}  for the transformed pulses are translated  for  the
physical laser pulses as 
\begin{equation}
  \label{eq:eqz:W:lin}
  \begin{pmatrix}
    C_{11} & \cdots & C_{1,N-1}\\
    C_{21} & \cdots & C_{2,N-1}\\
    \vdots & \vdots & \vdots \\
    C_{N-1,1} & \cdots & C_{N-1,N-1}
  \end{pmatrix}
  \begin{pmatrix}
    \Omega_1 \\
    \Omega_2 \\
    \vdots \\
    \Omega_{N-1}
  \end{pmatrix}
= 
\begin{pmatrix}
  f(t) \\
  f(t+\tau)\\
  0\\
  \vdots \\
  0
\end{pmatrix}
\end{equation}
which  is a  linear system  easily solvable  because  the coefficients
matrix  is  orthogonal.  We  stress  that  each pulse $\Omega_i$ is a
linear combination of pulses $f(t)$ and $f(t+\tau)$. Specific examples
will be given in  Section \ref{sec:numerics}, which is devoted to numerical 
solutions of the adiabatic evolution. We also notice that our derivation
remains valid for  more general parametrizations of transformed  pulses, e.g. 
$\tilde\Omega_1(t) =\Omega_0 (t) \sin(\phi(t))$, 
$\tilde\Omega_2(t) =\Omega_0 (t) \cos(\phi(t))$,
in which case the evolution would remain adiabatic whenever $|\phi'(t)/\Omega(t)| <<1$. 

\subsection{Case II: non-orthogonal initial and target states}

We consider the case in which the initial and target states are not
orthogonal:
\begin{equation}
  \label{eq:northo}
  \ps{\psi_f  }{ \psi_i} =  \cos \alpha  \neq 0  \qquad \alpha<  \pi/2~.
\end{equation}

We introduce a new basis $\{\ket{\phi_i}\}$ ($i=1,...,N-1$) for the
ground state subspace, with the first two basis vectors defined as
\begin{align}
  \label{eq:norto:lin:com}
  \ket{\phi_1} & = \ket{\psi_i}~, \\
  \ket{\phi_2} & = \frac{1}{\sin \alpha}\left[- \ket{\psi_f} + \cos\alpha \ket{\psi_i} \right]~.
%  \ket{\phi_3} & = \ket{\mathrm{comb. lin}} \\
 % \vdots &  \\
  % \ket{\phi_{N-1}} & = \ket{\mathrm{comb. lin}} 
\end{align}
and the remaining $N-3$ basis states determined by standard Gram- Schmidt 
orthogonalization procedure, so to complete the basis. In this new
basis, the final target state is expressed as
\begin{equation}
  \label{eq:not:orto}
  \ket{\psi_f} = \cos \alpha \ket{\phi_1} - \sin \alpha \ket{\phi_2}~.
\end{equation}
As for the case analyzed previously, we rewrite this as
\begin{equation}
\label{eq:norto:comb:lin:matr}
|\phi_i\rangle = \sum_j C_{ij} \ket{j} \qquad i,j=1,\ldots,N-1~,
\end{equation}
where the matrix $C$ is orthogonal. We proceed as before and introduce the new set of 
laser couplings $\tilde{\Omega}_i$ Eq. (\ref{eq:new:pulses}), and we parametrize them in 
terms of a total amplitude $\Omega$ Eq. (\ref{eq:totalomega}) and angles $\theta_i$ 
Eq. (\ref{eq:para:pulse}). The expressions for the dark states in terms of the parameters 
$\theta_i$, Eqs. \ref{eq:darkk} and in particular Eq. \ref{eq:dark1},  remain valid. 

Also in this case it is possible to reduce the system to an effective 
three-level $\Lambda$ system and implement the adiabatic evolution
from $|\psi_i \rangle$ to $|\psi_f \rangle$ as a standard STIRAP process.
The required form for the transformed pulses is: 
\begin{align}
%  \label{eq:cond:W:norto}
  \tilde{\Omega}_1 & \equiv f(t) \label{eq:cond:W:norto1}\,,\\
  \tilde{\Omega}_2 & \equiv \frac{1}{\tan \alpha} f(t) + f(t+\tau) \label{eq:cond:W:norto2}\,,\\
  \tilde{\Omega}_j & \equiv 0 \qquad j > 2 \label{eq:cond:W:norto3}\,.
\end{align}

As only $\tilde{\Omega}_1$, $\tilde{\Omega}_2$ are non-zero, the system is reduced 
to a three-level $\Lambda$ system, as in the case analyzed previously. Furthermore,
the specific choice of the pulse form, Eqs. (\ref{eq:cond:W:norto1},\ref{eq:cond:W:norto2}) 
leads to the transfer of the atomic system from $ \ket{\phi_1}$ to $ \cos \alpha \ket{\phi_1} - \sin \alpha \ket{\phi_2}$, i.e. from 
$\ket{\psi_i}$ to  $\ket{\psi_f}$. This can be shown by noticing that 
\begin{equation}
  \label{eq:form:theta:1}
  \tan \theta_1(t) = \frac{\tilde{\Omega}_1}{\tilde{\Omega}_2} =
  \frac{f(t)}{f(t+\tau)}
\rightarrow 
\begin{cases}
  0 & t \rightarrow -\infty\\
  \tan \alpha & t \rightarrow +\infty
\end{cases}  ~.
\end{equation}
Thus $\theta_1 (t\to -\infty)=0$ and $\theta_1 (t\to +\infty)=\alpha$ . Therefore,
the dark state $|\chi_1\rangle$ has the properties 
\begin{align}
  \label{eq:chi1:norto}
  \ket{\chi_1(-\infty) } &= \ket{ \psi_i }~, \\
  \ket{\chi_1(+\infty) } &= \ket{ \psi_f }~, 
\end{align}
and the adiabatic following of the dark state leads to the evolution of the system
from $\ket{ \psi_i } $ to $\ket{ \psi_f } $.

The physical laser pulses are obtained by solving the system of equations
(\ref{eq:eqz:W:lin}), and again each $\Omega_i$ is a linear combination of 
$f(t)$ and $f(t+\tau)$. 

\section{Numerical analysis}\label{sec:numerics}

In this Section we  prove the validity of our approach with numerical simulations. 
We numerically study the time-evolution of the atomic system to verify that our choice for 
the pulse sequence does indeed lead to adiabatic transfer from the initial to the  
target state. We consider a five-level system, with four ground states and one excited 
state. As already stressed, the procedure to identify the required pulse shape for the 
wanted transfer has minimal computational complexity, as it simply requires the inversion
of an orthogonal matrix, which corresponds to a transposition. Thus, the same procedure can be applied to larger atomic systems, 
with the same coupling structure,  without any computational difficulty.

%GSAs it will be shown, 
For fully adiabatic evolution, cf. Eq.~(\ref{Econdition_adiabatic}), 
the evolution of the atomic system does not involve populating the 
atomic excited state. 
Thus, we can study the time-evolution of the atomic system by solving the 
Schr\"odinger equation. In all numerical simulations 
presented here we will take the transformed pulses $\tilde{\Omega}_1 = f(t)$, 
$\tilde{\Omega}_2 = f(t+\tau)$ to have Gaussian shape
\begin{equation}
f(t)  = \Omega_0 \exp [-(t - t_0)^2/w^2],
\end{equation}
where $t_0$ is the pulse centre. The pulse delay $\tau$ between the delayed pulses is chosen so as to 
guarantee an overlap, and essential condition for the STIRAP process. The amplitude of the pulses 
$\Omega_0$, its width $w$ are chosen so to guarantee the adiabaticity of the process. We notice
that we chose the same amplitude $\Omega_0$ for the pulses for simplicity. However this is not 
a requirement for the STIRAP process in the effective three level $\Lambda$ system, and any 
combination of amplitudes for the two pulses $\tilde{\Omega_1}$, $\tilde{\Omega}_2$, such that 
the adiabaticity condition is satisfied, will lead to the correct implementation of the STIRAP process.

\begin{figure}[ht]
  \begin{center}
  \includegraphics[width=3in]{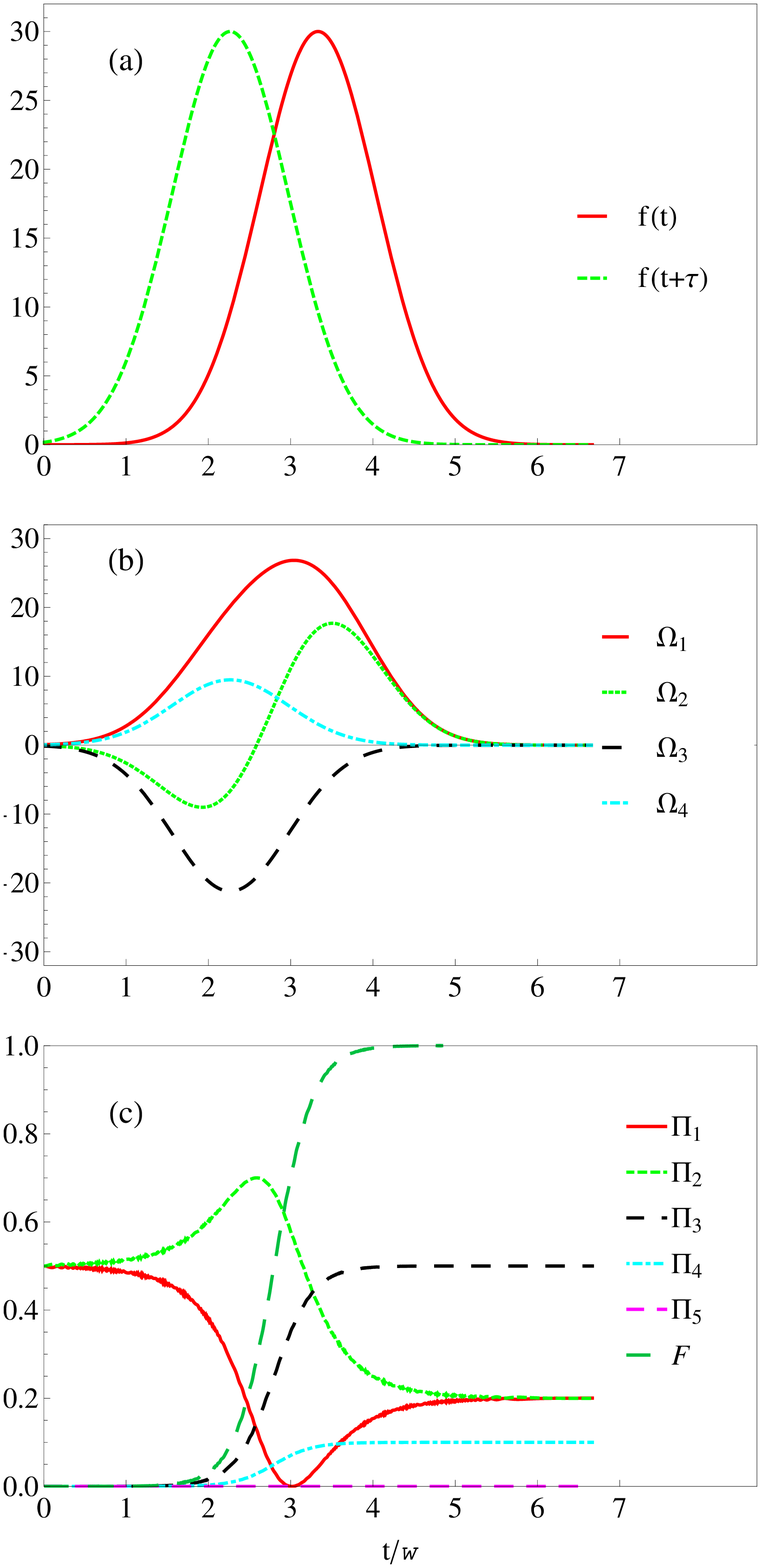}
\end{center}
\caption{(color online) Numerical solutions for the time-evolution of the 5-level system starting from  the  initial state
$|\psi_i\rangle = (|1\rangle+|2\rangle)/\sqrt{2}$. The pulses are determined, following the procedure
outlined in the text, to transfer the system into the state 
$|\psi_f\rangle = \sqrt{2/10}|1\rangle-\sqrt{2/10}|2\rangle - \sqrt{5/10}|3\rangle+\sqrt{1/10}|4\rangle $.  
(a) Pulse shapes for the Rabi frequencies $\tilde{\Omega}_1$, $ \tilde{\Omega}_2$  in the transformed 
basis $\{|\phi_i\rangle\}$.
(b) Pulse shapes for the Rabi frequencies $\Omega_i$  in the atomic basis $\{| i\rangle\}$.
(c) Population of the ground and excited states
%GS
in the atomic basis, and fidelity $F$ of preparation of the target state.
The parameters of the simulation are: $\Omega_0 =30$, $\tau = 160$, $w = 150$,  $t_0 = 500$.
}
\label{fig:numerics1}
\end{figure}

Figure \ref{fig:numerics1} shows  the solution of the time-dependent Schr\"odinger equation for 
the case of orthogonal initial and target states. Figure \ref{fig:numerics1}(a) reports the pulse 
shapes $f(t)$, $f(t+\tau)$ in the transformed basis, while in the initial basis the required laser pulses to obtain the wanted 
transfer are then determined via Eq. (\ref{eq:eqz:W:lin}), and are reported in Fig. \ref{fig:numerics1}(b).
We notice that the required relative sign between Rabi frequencies can experimentally be easily implemented by 
introducing a relative phase between the corresponding electric fields.
The resulting time-dependent populations $\Pi_i$ ($i=1-5$) of the different atomic states are reported in
Fig. \ref{fig:numerics1}(c), together with the fidelity of preparation of the wanted state 
$F=|\langle \psi_f|\psi(t)\rangle |^2$, where $|\psi(t)\rangle$ describes the state of the system
at time $t$. Our numerical results show that the fidelity approaches unity after the pulse sequence,
i.e. the system is effectively prepared in the wanted state $|\psi_f\rangle$.

\begin{figure}[ht]
\begin{center}
\includegraphics[width=3in]{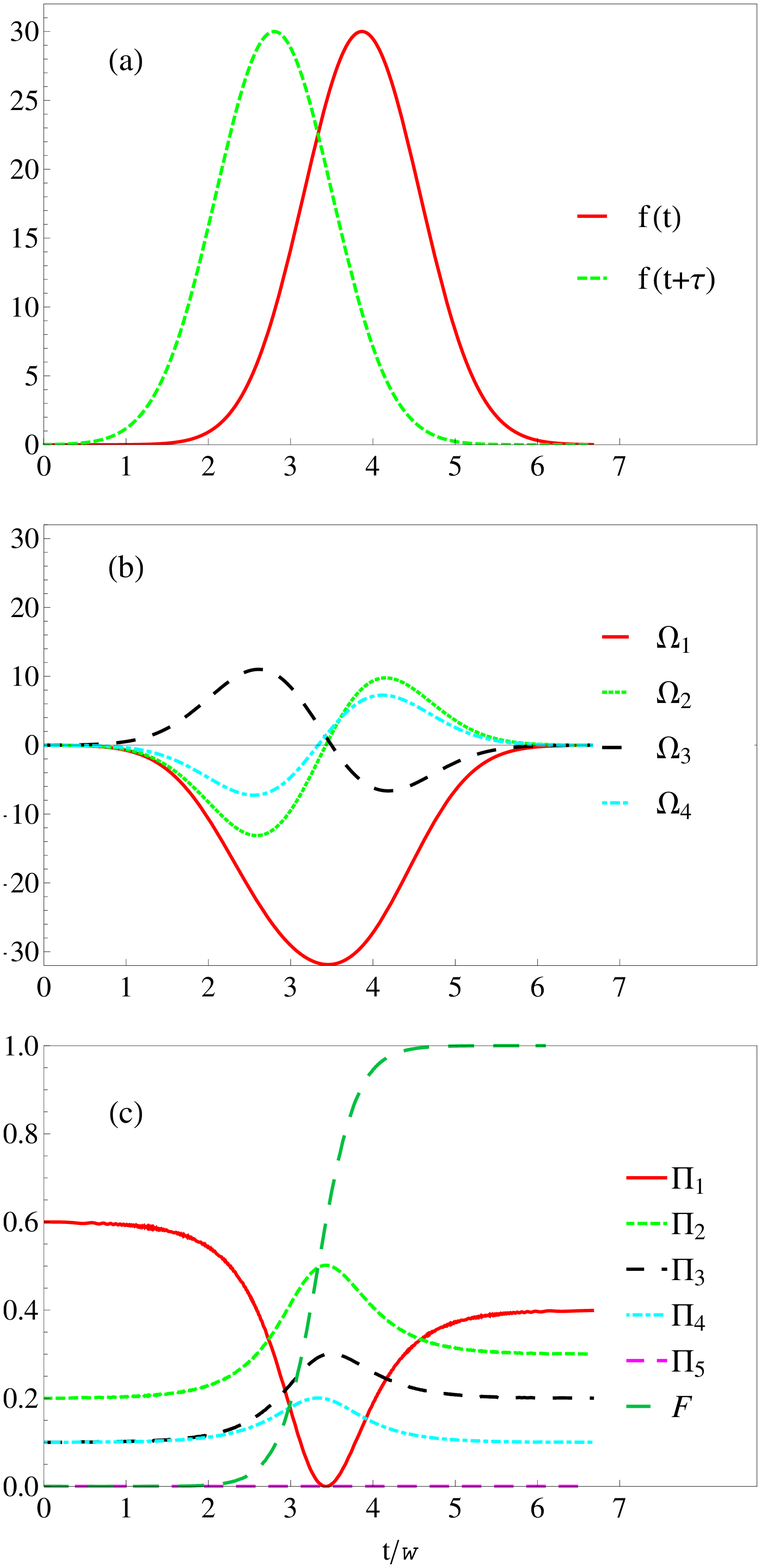}
\end{center}
\caption{(color online) Numerical solutions for the time-evolution of the 5-level system starting from  the  initial state
$|\psi_i\rangle = -\sqrt{3/5}|1\rangle+\sqrt{1/5}|2\rangle - \sqrt{1/10}|3\rangle+\sqrt{1/10}|4\rangle $. 
The pulses are determined, following the procedure outlined in the text, to transfer the system into the state 
$|\psi_f\rangle = \sqrt{2/5}|1\rangle+\sqrt{3/10}|2\rangle - \sqrt{1/5}|3\rangle+\sqrt{1/10}|4\rangle $.  
(a) Pulse shapes for the Rabi frequencies $\tilde{\Omega}_1$, $ \tilde{\Omega}_2$  in the transformed 
basis $\{|\phi_i\rangle\}$.
(b) Pulse shapes for the Rabi frequencies $\Omega_i$  in the bare atomic basis
basis $\{| i\rangle\}$.
(c) Population of the ground and excite states, and fidelity $F$ of preparation of the target state.
The parameters of the simulation are: $\Omega_0 =30$, $\tau = 160$, $w = 150$,  $t_0 = 500$.
}
\label{fig:numerics2}
\end{figure}

An analogous numerical analysis was also carried out for the case of non-orthogonal initial 
and target states. The procedure differs from the case analyzed previously only in the definition
of the transformed laser pulses. The non-orthogonality of the initial and target superpositions 
require a different transformation for $\tilde{\Omega}_2$, as given by Eq. (\ref{eq:cond:W:norto2}).
Our numerical results for this case, presented in  Fig. \ref{fig:numerics2}, confirm the validity of our 
approach: the process leads to a complete transfer from $|\psi_i\rangle$ to $|\psi_f\rangle$,
 without populating the excited state. Also in this case the amplitudes of the transformed fields 
were taken to be equal for simplicity. However, this is not a requirement for the STIRAP process, 
and arbitrary amplitudes can be used provided that they are large enough to guarantee the 
adiabaticity of the process.

\section{Conclusions}\label{sec:conclusions}

In this work we considered the problem of the implementation of Stimulated Raman Adiabatic 
Passage processes in degenerate systems, with a view to be able to steer
the system wave function from an arbitrary initial superposition to an 
arbitrary target superposition. We examined the case a $N$-level atomic system 
consisting of $ N-1$ ground states coupled to a common excited state by laser pulses. 
We analyzed the general case of initial and final superpositions belonging to the same 
manifold of states, and we cover also the case in which they are non-orthogonal.
We demonstrated that, for a given initial and target superposition,  it is always possible 
to choose the laser pulses so that in a transformed basis the system is reduced to an effective 
three-level $\Lambda$ system, and standard STIRAP applies. Our treatment leads to a  
simple strategy, with minimal computational complexity, which allows us to determine  the 
laser pulses shape required for the wanted adiabatic steering.

\acknowledgments

This work was supported by the Royal Society
and the DFG (BR 1528/7-1).

% Notice that choosing the pulses as 
% \begin{align}
%  \label{eq:cond:W:norto:gen}
%  \tilde{\Omega}_1 & \equiv a\, f(t) + b \,f(t+\tau)\\
%  \tilde{\Omega}_2 & \equiv c \, f(t) + d \,f(t+\tau) 
% \end{align}
% one has 
% \begin{equation}
 % \label{eq:form:theta:1:gen}
 % \tan \theta_1(t) = \frac{\tilde{\Omega}_1}{\tilde{\Omega}_2} =
 % \rightarrow 
%  \begin{cases}
 %   b/d & t \rightarrow -\infty\\
 %   a/c & t \rightarrow +\infty
% \end{cases}  
% \end{equation}

\end{document}